\documentclass[prl,preprint,nofootinbib]{revtex4}

\usepackage{graphicx}
\usepackage{subfigure}
\renewcommand{\vec}[1]{\ensuremath{\mathbf #1}}
\renewcommand{\matrix}[1]{\ensuremath{\mathrm #1}}

\newcommand{\bra}[1]{\ensuremath{\left \langle \left. #1
\right.  \right|  }}

\newcommand{\ket}[1]{\ensuremath{\left| \left. #1
\right. \right \rangle }}

\newcommand{\inner}[2]{\ensuremath{\left\langle \left. #1
\right.  \right| \left. #2 \right\rangle}}

\begin{document}

\title{Auxiliary density functionals: a new class of methods for
  efficient, stable density functional theory calculations}
\date{\today}
\author{P.J. Hasnip}\email{phil.hasnip@york.ac.uk}
\author{M.I.J. Probert}

\affiliation{Department of Physics, University of York, York YO10 5DD, U.K.}

\begin{abstract}
A new class of methods is introduced for solving the Kohn-Sham
equations of density functional theory, based on constructing a
mapping dynamically between the Kohn-Sham system and an auxiliary
system. The resulting auxiliary density functional equation is solved
implicitly for the density response, eliminating the instabilities
that arise in conventional techniques for simulations of large,
metallic or inhomogeneous systems. The auxiliary system is not
required to be fermionic, and an example bosonic auxiliary density functional
is presented which captures the key aspects of the fermionic Kohn-Sham
behaviour. This bosonic auxiliary scheme is shown to provide good
performance for a range of bulk materials, and a substantial
improvement in the scaling of the calculation with system size for a
variety of simulation systems.
\end{abstract}


\maketitle  

Density functional theory (DFT)\cite{Hohenberg-Kohn} has been used
successfully to study a vast range of chemicals and materials,
predicting everything from crystal structures and bulk moduli, to
XANES and NMR spectra\cite{Castep-spectroscopy}. Its widespread
success across the physical sciences has led to the creation of entire
computational sub-disciplines in fields as diverse as chemistry,
physics, materials science, engineering and Earth sciences\cite{DFT-success}, as well as
increasing application to the biological sciences. Furthermore, DFT
simulations have become an essential tool for many experimental 
chemical and materials studies, where they are used to facilitate the
interpretation of experimental data and guide experimental design.

In order for a materials simulation to have \emph{predictive} power,
it must take into account the quantum mechanical nature of the
materials' constituents, in particular those of the valence electrons
whose behaviour governs the mechanical, electronic and chemical
properties of matter. In principle the electronic states could be
computed by solving the Schr\"odinger equation to find the many-body
wavefunction for the electrons, but the computational complexity of
this approach renders it impractical for realistic systems. DFT
provides a solution to this problem by showing that only the
electronic density is necessary to compute a system's groundstate
properties, not the full many-body wavefunction of a system.

DFT is an exact reformulation of quantum mechanics, founded on the
Hohenberg-Kohn theorem which states that the groundstate energy of a
quantum system is a unique functional of the groundstate
density\cite{Hohenberg-Kohn}\cite{Levy}. Thus in principle the
groundstate energy and density of any system may be computed by
minimizing the energy with respect to the density, without the need to
determine the many-body wavefunction. Unfortunately the form of the
density functional is not known at present.

Kohn and Sham introduced a practical DFT framework\cite{Kohn-Sham} by
mapping the ground state of the interacting $N$-body fermionic system
 onto that of an auxiliary
\emph{non-interacting} fermionic system. This mapping transforms the
groundstate solution of the original $N$-body Schr\"odinger equation
into the $N$ lowest energy solutions of a 
single-particle Schr\"odinger equation (the Kohn-Sham equation), for which the kinetic, Coulomb and external potential
functionals have a known, simple form. The Hohenberg-Kohn theorem
guarantees that the groundstate properties of this auxiliary system
will match those of the original many-body system, provided they have
the same groundstate density; in order for this to be the case, an
additional functional is included in the auxiliary Kohn-Sham Hamiltonian, referred to as the
\emph{exchange-correlation} functional. The exact form of this exchange-correlation
functional is not known, but as its contribution to the total energy
is small, even relatively crude approximations can yield useful
results. In this work we extend the mapping of Kohn and Sham to
a broader class of auxiliary systems, and show how this may be used to
improve the performance and stability of Kohn-Sham DFT simulations.


Central to any conventional DFT simulation is the solution of the
single-particle Kohn-Sham equation. The construction of
the charge density is time-consuming and so it is common to solve the Kohn-Sham
equation for a given input density $n_\mathrm{in}(\vec{r})$:
\begin{equation}\label{eq:ks}
\left(-\frac{1}{2}\nabla^2 + \hat{V}_\mathrm{Hxc}[n_\mathrm{in}] + \hat{V}_\mathrm{ext}\right)\psi_i(\vec{r}) = \epsilon_i\psi_i(\vec{r})
\end{equation}
where $V_\mathrm{Hxc}[n]$ is the Hartree-exchange-correlation potential,
$V_\mathrm{ext}$ is the external potential and $\psi_i(\vec{r})$ is the
wavefunction of the $i$th eigenstate (particle) at position
$\vec{r}$. Clearly in general $n_\mathrm{in}(\vec{r})$ is not the same as the density
computed from the wavefunctions, $n_\mathrm{out}(\vec{r})$, defined as:
\begin{equation}\label{eq:density}
n_\mathrm{out}(\vec{r}) = \sum_j f_j\left\vert \psi_j(\vec{r}) \right\vert^2,
\end{equation}
where $\{f_j\}$ are the eigenstate occupation numbers. A separate algorithm is
employed to evolve $n_\mathrm{in}(\vec{r})$ towards the groundstate density,
and it is this algorithm which is the focus of this work. Regardless
of the method used, once  $n_\mathrm{in}(\vec{r})$ has reached the groundstate
it must satisfy the so-called self-consistency condition
\begin{eqnarray}
n_\mathrm{in}(\vec{r})=n_\mathrm{out}(\vec{r})\label{eq:selfcon}
\Rightarrow \hat{V}_\mathrm{Hxc}[n_\mathrm{in}] = \hat{V}_\mathrm{Hxc}[n_\mathrm{out}];
\end{eqnarray}
for this reason, these methods are known generally as
\emph{self-consistent field} (SCF) methods. Whenever $n_\mathrm{in}(\vec{r})$ is
updated, Equation~\ref{eq:ks} must be solved to generate the new
$\{\psi_i(\vec{r})\}$, and hence $n_\mathrm{out}(\vec{r})$. It is convenient
to measure the error in self-consistency at iteration $k$ by the
density residual $R^{(k)}$, defined as
\begin{equation}
R^{(k)}\left(\vec{r}\right)=\left(n^{(k)}_\mathrm{out}\left(\vec{r}\right)-n^{(k)}_\mathrm{in}\left(\vec{r}\right)\right).
\end{equation}
When Equations~\ref{eq:ks} and \ref{eq:selfcon} are satisfied
simultaneously then $R^{(k)}\left(\vec{r}\right)=0$ and the system has
reached the ground state.

The ability of this method to solve the Kohn-Sham equations
depends critically on the algorithm used to generate
$n^{(k)}_\mathrm{in}(\vec{r})$, the new input density for iteration $k$, from
$n^{(k-1)}_\mathrm{in}(\vec{r})$ and $R^{(k-1)}(\vec{r})$. In fact all the common
approaches, known as density mixing methods, are based on a linear
combination of many previous density residuals,
\begin{equation}
n^{(k)}_\mathrm{in}\left(\vec{r}\right)=n^{(k-1)}_\mathrm{in}\left(\vec{r}\right)+\sum_{m<k} \alpha_m R^{(m)}\left(\vec{r}\right),
\end{equation}
where $\{\alpha_m\}$ are real coefficients in the interval [0,1].

The popular density mixing methods of Pulay\cite{Pulay} and
Broyden\cite{Broyden} work by linearising the response of the
Kohn-Sham system, assuming that
\begin{equation}\label{eigendensity}
n^{(k)}_\mathrm{out}(\vec{r}) = \hat{M}n^{(k)}_\mathrm{in}(\vec{r})
\end{equation}
where $\hat{M}$ is a linear operator (though not necessarily
Hermitian). In general an approximation to $\hat{M}$ is constructed
successively by a low-rank update at each iteration, e.g.
\begin{equation}\label{eq:low_rank}
\hat{M}^{(k)} = \sum_{i<k}\sum_{j<k} \ket{n_\mathrm{out}^{(i)}}\matrix{M}^{-1}_{ij}\bra{n_\mathrm{in}^{(j)}},
\end{equation}
where $\matrix{M}$ is a $k \times k$ subspace matrix defined as
\begin{equation}
\matrix{M}_{ij} =  \inner{n_\mathrm{in}^{(i)}}{n_\mathrm{in}^{(j)}}.
\end{equation}

Thus these methods build up an improved approximation to $\hat{M}$
using the information obtained from each iteration. If $\hat{M}$ is a
good representation of the true Kohn-Sham response, then the
eigendensity of $\hat{M}$ (as defined by Equation~\ref{eigendensity})
is a good approximation to the self-consistent density of the
Kohn-Sham system.

As the size of simulation system is increased, the convergence of
simple density mixing schemes becomes poorer and these algorithms
often diverge for large simulation systems. It is well-known\cite{HIJ}
that the root cause of this divergence is the behaviour of the Hartree
potential $V_\mathrm{H}$, which in reciprocal space may be expressed as
\begin{equation}
V_\mathrm{H}\left(\vec{G}\right) = \frac{4\pi \tilde{n}_\mathrm{in}\left(\vec{G}\right)}{\Omega\left\vert\vec{G}\right\vert^2},
\end{equation}
where $\tilde{n}_\mathrm{in}$ is the Fourier transform of
${n}_\mathrm{in}$, \vec{G} is a reciprocal lattice vector and $\Omega$ is the
volume of the simulation cell. As the size of simulation cell is
increased, the size of the smallest \vec{G}-vector decreases and a
small error in $n_\mathrm{in}$ can lead to a large error
$V_\mathrm{H}[n_\mathrm{in}]$; this in turn leads to a large error in
$n_\mathrm{out}$ and the iterative scheme will diverge. This instability with
respect to small changes in $n$ is often referred to as a `sloshing instability'.

A simple preconditioner was introduced by Kerker\cite{Kerker} in an
attempt to correct for this ill-behaviour, based on the iterative
scheme proposed for jellium by Manninen et al.\cite{Manninen}, and this can be
incorporated into the density mixing schemes:
\begin{equation}
\tilde{n}^{(k)}_\mathrm{in}(\vec{G})=\tilde{n}^{(k-1)}_\mathrm{in}(\vec{G})+\sum_{m<k} \alpha_m \left(\frac{\left\vert\vec{G}\right\vert^2}{\vert\vec{G}\vert^2+G_0^2}\right)\tilde{R}^{(m)}(\vec{G})
\end{equation}
where $G_0$ is a characteristic reciprocal length. The residual
components with $\left\vert\vec{G}\right\vert << G_0$
are multiplied by $\left\vert\vec{G}\right\vert^2$, approximately
  correcting for the $\frac{1}{\left\vert\vec{G}\right\vert^2}$
  divergence in $V_\mathrm{H}$, whereas the components of large wavevectors are unchanged.

By combining this preconditioner with a density mixing method,
reasonable convergence can be attained for many moderately sized
simulation cells. The characteristic reciprocal length $G_0$ is a
parameter of the scheme, but it is usually found that $G_0=1.5\AA^{-1}$\ is
an acceptable approximation for bulk systems\cite{Kresse}.

Preconditioned density mixing methods converge reasonably well for
many bulk systems, but often struggle with large, metallic or
anisotropic simulation systems. There are three main reasons for this:
firstly, the Kerker preconditioner is isotropic, with a scalar
characteristic reciprocal length $G_0$; secondly the Kerker
preconditioner ignores other contributions to the response, in
particular that due to the exchange-correlation functional; finally
the linearisation of the charge density response is only valid near
the groundstate, so the methods require a good initial input density.

In principle the exact dielectric response could be calculated from
the Kohn-Sham states, at least within the linear regime. More
practically, the response tensor could be computed for the subset
containing the smallest \vec{G}-vectors, a method advocated by Ho
\emph{et al}.\cite{HIJ}. Unfortunately this method is unstable when
the non-linear response is significant, and furthermore the
computational cost of the dielectric tensor becomes prohibitive as the
simulation size increases. This latter issue has been addressed
recently by Krotscheck and Liebrecht\cite{Krotscheck}, who use
linear-response perturbation theory to account for the dielectric
response.

The key to making all such schemes tractable is in realising that only a
single eigendensity is required, and so Equation~\ref{eigendensity}
may be solved efficiently using iterative methods. Such methods do not
require the explicit construction of the response tensor, only the
application of it to successive trial eigendensities\cite{Krotscheck}. In this spirit, Raczkowski \emph{et al}.\cite{TF-mixing} proposed replacing the Kerker
preconditioner with one based on the Thomas-Fermi-von Weizs\"acker
(TFW) functional. In this latter scheme the dielectric response is
approximated by a modified TFW system, and the eigendensity computed using
an iterative scheme. This eigendensity is then used as the input to a
conventional Pulay mixing method.

Whilst this TFW-based scheme avoids the explicit construction of the
dielectric tensor, it suffers from two major drawbacks: firstly by
implementing the method merely as a preconditioner, much of the
advantages of the response calculation are lost by the subsequent
Pulay or Broyden mixing scheme; secondly the iterative preconditioner
they proposed is not itself stable, and suffers from unphysical
run-away solutions. Nevertheless the central concept of using an
auxiliary functional to model the response is sound. 

In this work a new class of SCF methods is introduced, based on the
use of auxiliary density functionals, which do not require either
density-mixing or an explicit dielectric tensor. A unified framework
for constructing such auxiliary functionals is proposed which
avoids the separation of the preconditioner and the general
dielectic-modelling method and, crucially, includes the full
non-linear response due to  $V_\mathrm{Hxc}$.

In this approach, the Kohn-Sham system is itself modelled using an auxiliary
system, with a corresponding auxiliary density functional,
$\hat{H}_0\left[n\right]$. This mapping from the Kohn-Sham system to
the auxiliary system is exactly analogous to that between the
Kohn-Sham system and the original many-body interacting system; indeed
the existence of an exact auxiliary mapping is guaranteed by the
Hohenberg-Kohn theorem.

In keeping with conventional SCF methods, the Kohn-Sham wavefunctions
are relaxed for a given input density, $n_\mathrm{in}$. Once the
ground state has been obtained for this input density, the Kohn-Sham
wavefunctions are used to generate a new density, the output density
$n_\mathrm{out}$. The goal of the SCF method is to find the input
charge density which is invariant under this process, i.e. to find
$n_\mathrm{in}$ such that $n_\mathrm{out} = n_\mathrm{in}$
(Equation~\ref{eq:selfcon}). As with the Pulay and Broyden schemes,
the Kohn-Sham response is modelled to provide a good input density
$n_\mathrm{in}$ for the next iteration, but rather than linearising
the full response and using matrix methods, in this approach the
Kohn-Sham system is modelled using an auxiliary system, with an
auxiliary density functional, $\hat{H}_0$, which crucially includes
the full, non-linear $V_\mathrm{Hxc}$. This mapping from the Kohn-Sham
system to the auxiliary system is exactly analogous to that between
the Kohn-Sham system and the original many-body interacting system;
indeed the existence of an exact auxiliary mapping is guaranteed by
the Hohenberg-Kohn theorem.

There are many possible choices for the model functional, each
corresponding to a different auxiliary system. As an exemplar, we here
introduce one of the simplest such choices: a mapping of the fermionic
Kohn-Sham system onto an auxiliary system of bosons. This bosonic mapping leads to the simple Kohn-Sham-like functional for
the auxiliary system
\begin{equation}
\hat{H}_0\left[n\right] = \left( -\frac{1}{2}\nabla^2 + \hat{V}_\mathrm{Hxc}\left[n\right] + \hat{V}_\mathrm{ext} \right).
\end{equation}
A key feature of this auxiliary density functional is that it has the
correct behaviour with respect to $V_\mathrm{Hxc}$ and the local
$V_\mathrm{ext}$, which is required to suppress sloshing instabilities. The
kinetic and any non-local $V_\mathrm{ext}$ contributions are approximations
to those of the Kohn-Sham system, but are correct in the limit of a
simulation system with a single state (the single-orbital
approximation of Hodgson \emph{et al.}\cite{Hodgson}); in order for the auxiliary
bosonic system to model the Kohn-Sham system accurately it is
necessary to improve this kinetic-non-local approximation dynamically
over the course of a simulation.

Each iteration $k$ of the Kohn-Sham solver (Equation~\ref{eq:ks})
generates a new $n^{(k)}_\mathrm{in}$ and $n^{(k)}_\mathrm{out}$ pair, which yield
information on the non-self-consistent response of the fermionic
Kohn-Sham system. In order to generate an accurate mapping between the
Kohn-Sham and 
auxiliary systems, the
auxiliary bosonic system must yield the same density
response as the Kohn-Sham system when the input density is fixed; 
since in general this will not be the case, a correction operator
$\hat{P}$ is added to the auxiliary Hamiltonian such that at iteration $k$
\begin{equation}\label{modified} 
\left(\hat{H}_0\left[n_\mathrm{in}^{(k)}\right]+ \hat{P}\right)\sqrt{n^{(k)}_\mathrm{out}}= E_F^{(k)}\sqrt{n^{(k)}_\mathrm{out}}.  
\end{equation} 
where $E_F$ is the Fermi energy. In order for Equation \ref{modified} to be satisfied, 
\begin{equation}\label{def:P}
\hat{P}\sqrt{n^{(k)}_\mathrm{out}} = \left(E_F^{(k)} -
\hat{H}_0\left[n_\mathrm{in}^{(k)}\right]\right)\sqrt{n^{(k)}_\mathrm{out}}.
\end{equation}

In general $\hat{P}=\hat{P}\left[n\right]$, with a corresponding
energy correction $E_P\left[n\right]$. This functional is unknown, but
a Taylor expansion of $E_P$ shows that the first-order variation with
$n$ may be captured by a fixed operator $\hat{P}$. In contrast to
density-mixing methods, this linearisation applies only to the
correction energy; in particular the non-linear exchange-correlation
response is modelled exactly.

There is considerable freedom in the choice of
$\hat{P}$, but the success of the Pulay and Broyden schemes suggests a low-rank
approximation may be productive, for
example the Hermitian form
\begin{equation}
\hat{P}^{(k)} = \frac{\ket{q_k}\bra{q_k}}{\inner{q_k}{p_k}}\label{def:rank-1},
\end{equation}
where 
\begin{eqnarray}
p_k&=&\sqrt{n^{(k)}_\mathrm{out}}\label{def:p}\\
q_k&=&\left(E_F^{(k)} - \hat{H}_0\left[n_\mathrm{in}^{(k)}\right]\right)\sqrt{n^{(k)}_\mathrm{out}}.\label{def:q}
\end{eqnarray}

%

This form for $\hat{P}^{(k)}$ is well-defined provided
$\inner{q_k}{p_k}\neq0$ is 
real (which follows naturally from the definitions in equations~\ref{def:P},
\ref{def:p} and \ref{def:q}) and non-zero; in the case
$\inner{q_k}{p_k}=0$, Equation~\ref{def:P} is satisfied by $\hat{P}=0$.

Once a suitable auxiliary system has been chosen and the correction
operator $\hat{P}$ determined, the Kohn-Sham
system is modelled using the Schr\"odinger-like equation for the
square-root of the density
\begin{equation}\label{adft}
\left( \hat{H}_0\left[n\right]+ \hat{P}^{(k)}\right)\sqrt{n} = E_F^{(k)}\sqrt{n},
\end{equation}
The (self-consistent) solution to
Equation~\ref{adft} is the ground state density for the auxiliary
system, $n_\mathrm{aux}$, and an estimate of the ground state density of the
Kohn-Sham system. This density is thus used as the new input density
for the Kohn-Sham solver (Equation~\ref{eq:ks}),
i.e. $n_\mathrm{in}^{(k+1)}=n_\mathrm{aux}$.

As the bosons may all occupy the same state simultaneously, there is
no need to compute and orthonormalise large
numbers of states for the auxiliary system; $n\left(\vec{r}\right)$ is
the charge density of the entire auxiliary system,
with no need for a further summation over bands or Brillouin zone
sampling points. With an appropriate preconditioner\cite{precon} the computational effort to solve Equation
\ref{adft} scales approximately linearly with system size, and for
large simulation systems its solution time is insignificant compared
to solving Equation~\ref{eq:ks}, despite the similarity between the
two. 

This auxiliary density functional theory (ADFT) scheme has been
implemented in a development version of the CASTEP plane-wave
pseudopotential computer program\cite{CASTEP}. The charge density is
set initially to a superposition of pseudoatomic charge densities, and
the present scheme compared to the existing Kerker, Pulay and Broyden
density mixing methods, the latter two methods using a Kerker
preconditioner. To illustrate the power of the ADFT method, results
for a memory-less approximation to $\hat{P}$ are reported, whereby
$\hat{P}^{(k)}$ is determined using only information from iteration
$k$; in contrast the Pulay and Broyden calculations reported here use
information from all previous iterations.

The initial tests consisted of applying the bosonic ADFT method to a
variety of simple materials with different electronic properties,
spanning the range from metals to wide band-gap insulators and with
s-, p- and d-states. Performance was measured by comparing the number
of iterations required for convergence with that of a
conventional Pulay density mixing method; in each case the calculation
was deemed to have converged when the energy was less than 2
$\mu$eV/atom from the (precomputed) groundstate energy. For all of
these tests the Pulay and bosonic ADFT methods showed comparable
performance (see Table \ref{bulk_results}); since existing density
mixing methods are known to perform well for small, isotropic systems
such as these
the benefit of the ADFT scheme is expected to be for larger,
anisotropic simulations such as the surface, interface and defect
calculations that characterize much of nanomaterials, surface science
and atomic growth research.

\begin{table}
\begin{tabular}{l|ccccccc}
& \multicolumn{7}{c}{Material}\\
Method & K & Al & Au & Si & MgO & GaN & graphite \\
\hline
Pulay & 6 & 5 & 6 &  6  & 7 & 8 & 6 \\
ADFT & 5 & 5 & 6 & 5 & 5 & 6 & 5 
\end{tabular}
\caption{A comparison of the number of SCF iterations required to converge
  to the ground state for a Pulay mixing scheme and the bosonic auxiliary DFT scheme for a variety of simple bulk materials.}\label{bulk_results}
\end{table}

\textbf{Magnesium oxide (MgO)} is a wide-band-gap insulator with
potential applications in future CMOS devices. It has been widely
studied with \emph{ab initio} methods, including recent studies of
electron tunnelling\cite{Keith_MgO}, layer-by-layer film
growth\cite{MgO_PRL}, and its use as an interlayer in stabilising
heterostructures\cite{MgO_interlayer}. MgO is also a prototypical
polar oxide, and forms an ideal testing ground for the
application of the ADFT scheme to ionically bonded systems. 

\begin{figure}
\begin{center}
\includegraphics[width=0.45\textwidth]{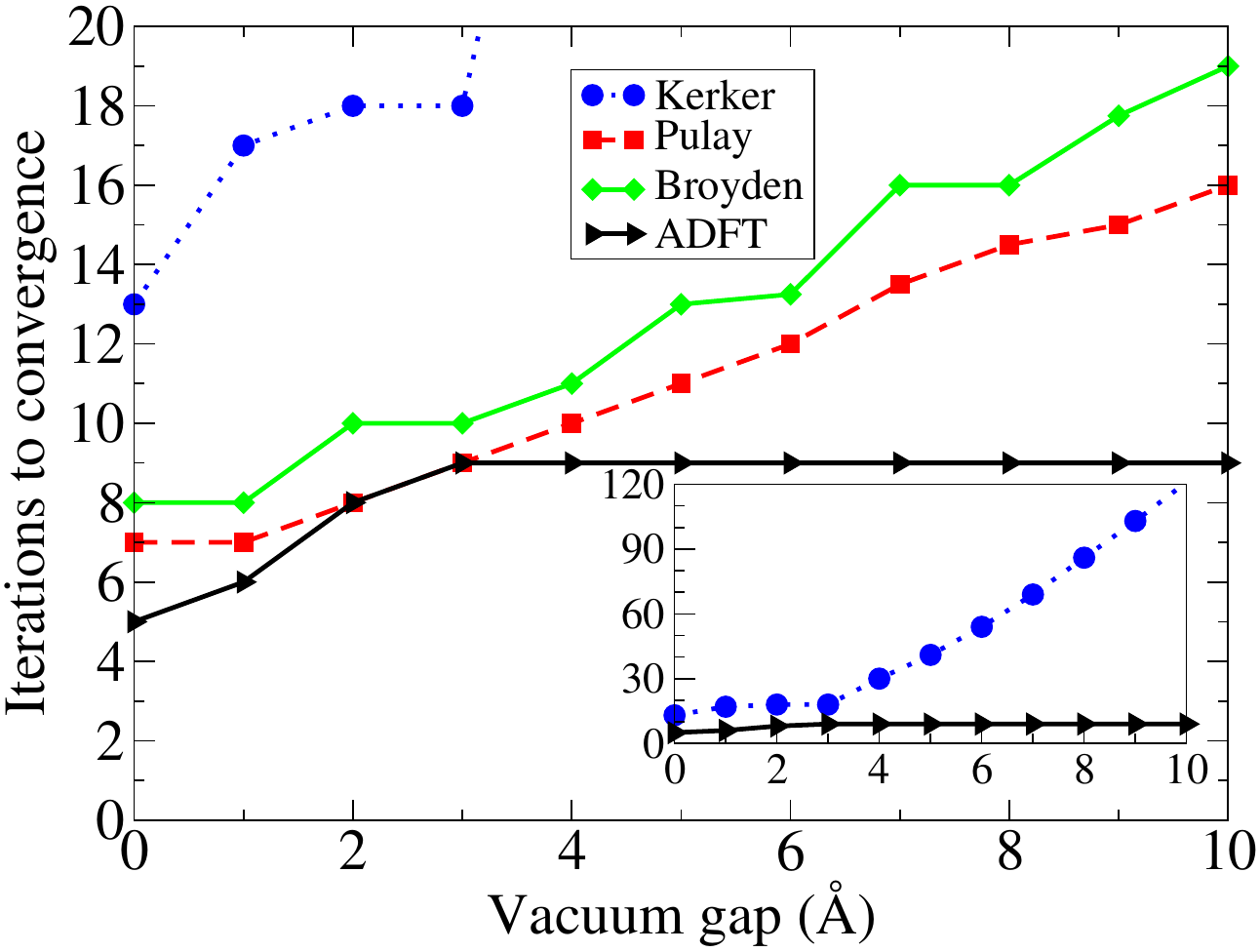}
\caption{(Color online) Convergence of the bosonic ADFT
  scheme compared to the Pulay, Broyden and Kerker (inset) density mixing
  methods for a 2D slab of MgO as the vacuum gap between periodic slab
  images is varied.  }\label{Fig:MgO}
\end{center}
\end{figure}

Bulk MgO has a rocksalt structure and conventional density mixing
algorithms perform well (see Table~\ref{bulk_results}). In order to
simulate thin films the bulk crystal was cleaved along (001) and the
resulting slabs separated from their periodic image by a vacuum
region. The convergence behaviour of the ADFT and density mixing
methods was then studied as a function of the vacuum separation (Figure~\ref{Fig:MgO}).

The strength of the ADFT scheme is apparent as the vacuum gap is increased: the
performance of the density mixing schemes degrades considerably, with
the number of iterations growing approximately linearly with system
size, whereas the ADFT performs equally well for all system sizes. The
consistent ADFT performance is because the response of the Kohn-Sham
system is dominated by the Hartree-exchange-correlation and local
external potential contributions, and these contributions are captured
exactly by the auxiliary bosonic system.

\textbf{Graphene} has a wide variety of potential applications\cite{Graphene},
from flexible display technologies to transistors. It is comprised of
a single 2D sheet of carbon hexagons and many of its possible
applications arise from its unusual bandstructure: graphene is a
semi-metal, with a Dirac cone at the Fermi level. For this reason the electronic structure of graphene and
graphene-derivatives are of immense interest, but the computational
time required by DFT algorithms has led to researchers using quicker
tight-binding models for larger
geometries\cite{Yvette1}\cite{Yvette2}.

\begin{figure}
\begin{center}
\includegraphics[width=0.45\textwidth]{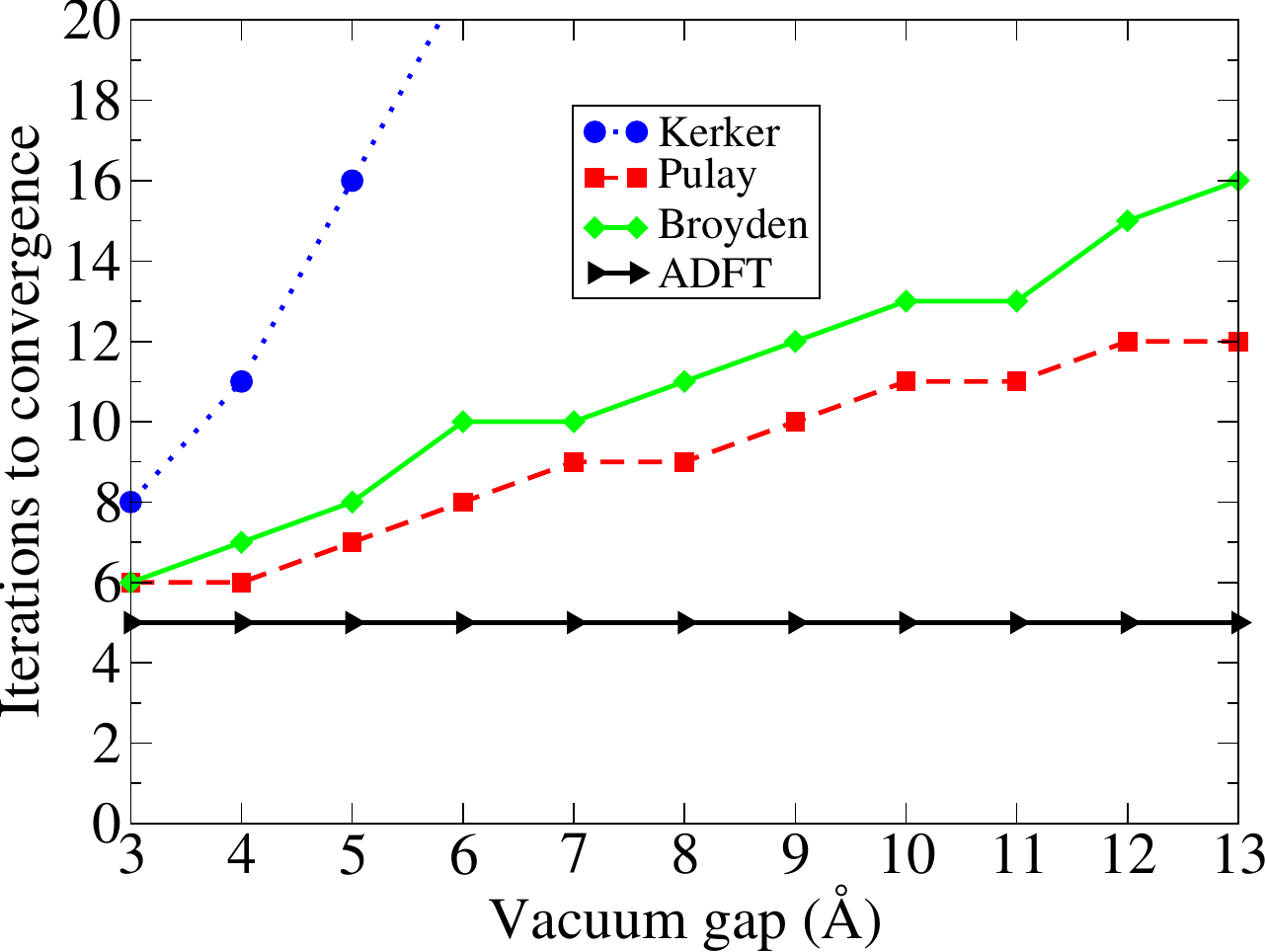}
\caption{(Color online) Performance of the bosonic ADFT scheme for graphene, as a
  function of the spacing between periodic images in the perpendicular
  direction. The number of iterations taken to converge is
  approximately constant, whereas the iteration count for conventional
  density mixing methods grows.}\label{Fig:graphene}
\end{center}
\end{figure}

To simulate graphene a primitive cell was constructed and the
convergence behaviour of the various methods was studied as the
inter-layer separation was increased from 3\AA\ (comparable to the
layer separation in graphite) to 13\AA\ by which point each layer is
decoupled from its periodic images. As the system size increases the
iterations required for conventional density mixing method grows
linearly, whereas the ADFT scheme performs consistently well (Figure
\ref{Fig:graphene}).

\textbf{Gold nanoparticles} have a wide variety of medical
applications as well as an increasing array of applications in
electron microscopy and nanomaterials research. Recent DFT simulations
have focused on their high catalytic activity, in particular for
oxidation of carbon monoxide\cite{Keith_Au}.

\begin{figure}
\begin{center}
\includegraphics[width=0.45\textwidth]{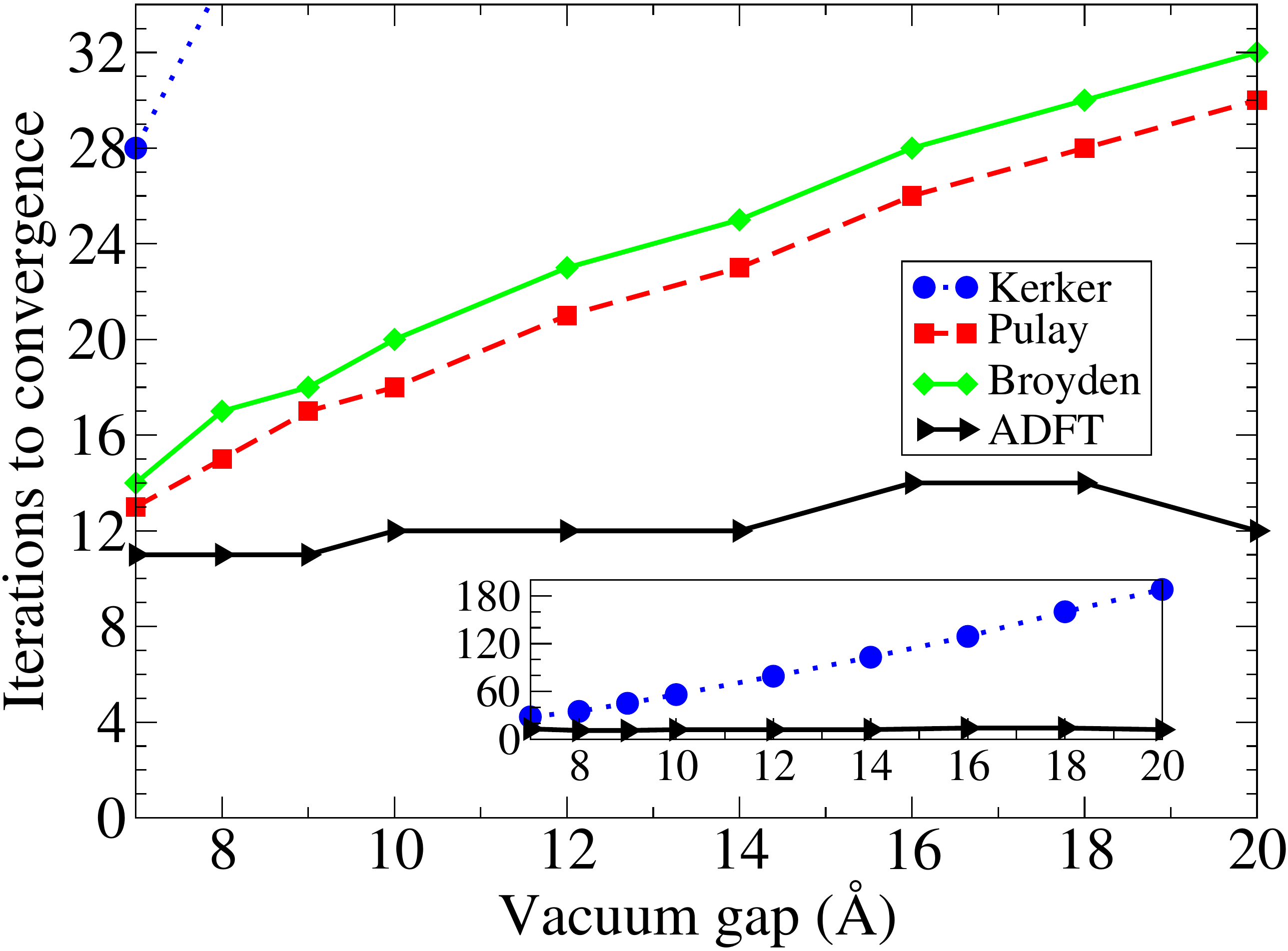}
\caption{(Color online) Performance of the bosonic ADFT scheme
  compared to the Pulay, Broyden and Kerker (inset) density mixing
  methods for a gold nanoparticle, as a
  function of the spacing between periodic images.}\label{Fig:Au}
\end{center}
\end{figure}

To simulate a simple gold nanoparticle a primitive cell was
constructed containing a gold tetramer, and the inter-particle separation
was increased from 7\AA\ to 20\AA\ at which size each particle is decoupled
from its periodic images. The convergence behaviour of the various
methods can be seen in Figure~\ref{Fig:Au}. As before, the performance
of the ADFT method is consistent across the range of cell sizes
whereas that of the conventional density mixing methods degrades as
the vacuum separation is increased, with a linear increase in the
number of iterations with system size.

In conclusion, an auxiliary density functional theory framework is
introduced that maps the fermionic Kohn-Sham system onto an auxiliary
system, which is not constrained to be fermionic. This mapping is
constructed dynamically during an ordinary Kohn-Sham DFT calculation,
and can be used to accelerate the convergence of SCF calculations
considerably. The power of this ADFT framework is illustrated using a
bosonic auxiliary system and the performance shown to be invariant
with the size of the simulation cell, eliminating the `sloshing'
instabilities common in other techniques and removing the linear
scaling of iteration count with system size. The method has been
implemented in a development version of the CASTEP computer program,
and demonstrates significantly superior performance to existing
methods for a variety of systems from a wide-band-gap polar oxide thin
film, to a 2D Dirac semi-metal and an isolated nanocluster.

The authors would like to thank M.~J.~Verstraete, M.~Stankovski, H. Ness,
J.~D.~Ramsden, M.~J.~P.~Hodgson, K.~Burke, M.~Levy and especially
R.~W.~Godby for many useful discussions. Financial support from the
EPSRC and the United Kingdom Car-Parrinello Consortium is acknowledged
with gratitude.


\end{document}